\documentclass[11pt,a4paper]{article}
\usepackage[utf8]{inputenc}
\usepackage[margin=25mm]{geometry}
\usepackage[english]{babel}
\usepackage[numbers]{natbib}

\usepackage{csquotes}
\usepackage{color}
\usepackage{hyperref}
\usepackage{url}
\usepackage{breakurl}

\title{
When Curation Becomes Creation:
Algorithms, Microcontent, and the Vanishing Distinction between Platforms and Creators
}
\author{Liu Leqi, Dylan Hadfield-Menell, Zachary C. Lipton \\
\href{mailto:leqi@cs.cmu.edu}{\nolinkurl{leqi@cs.cmu.edu}}, \href{mailto:dylanhm@mit.edu}{\nolinkurl{dhm@csail.mit.edu}}, \href{mailto:zlipton@cmu.edu}{\nolinkurl{zlipton@cmu.edu}}
}
\begin{document}
\maketitle

Ever since social activity on the Internet began migrating from the wilds of the open web to the walled gardens erected by so-called \emph{platforms} (think Myspace, Facebook, Twitter, YouTube, or TikTok), debates have raged about the responsibilities that these platforms ought to bear. And yet, despite intense scrutiny from the news media and grassroots movements of outraged users, platforms continue to operate, from a legal standpoint, on the friendliest terms. 

You might say that today's platforms enjoy a ``have your cake, eat it too, and here's a side of ice cream'' deal. They simultaneously benefit from: (1) broad discretion to organize (and censor) content however they choose; (2) powerful algorithms for curating a practically limitless supply of user-posted microcontent according to whatever ends they wish; and (3) absolution from almost any liability associated with that content.

This favorable regulatory environment results from the current legal framework, which distinguishes between intermediaries (e.g., platforms) and content providers. This distinction is ill-adapted to the modern social media landscape, where platforms deploy powerful data-driven algorithms (so-called AI) to play an increasingly active role in shaping what people see and where users supply disconnected bits of raw content (tweets, photos, etc.) as fodder. 

Specifically, under Section 230 of the Telecommunications Act of 1996, ``interactive computer services'' are shielded from liability for information produced by ``information content providers.'' While this provision was originally intended to protect telecommunications companies and Internet service providers from liability for content that merely passed through their plumbing~\citep{sec230history}, the designation now shelters services such as Facebook, Twitter, and YouTube, which actively shape user experiences. 

Excepting obligations to take down specific categories of content (e.g., child pornography and copyright violations), today's platforms have license to monetize whatever content they like, moderate if and when it aligns with their corporate objectives, and curate their content however they wish. 

\subsection*{ANTECEDENTS IN MODERATION} 
In his 2018 book, Custodians of the Internet~\cite{gillespie2018custodians}, Tarleton Gillespie examines platforms through the lens of content moderation, calling into focus an apparent contradiction: Platforms constantly do (and arguably, must) wade into the normative, making political decisions about what content to allow; and yet they operate absent responsibility on account of their purported \emph{neutrality}. 

Throughout, Gillespie is even-handed, expressing sympathy for platforms' predicament. They must moderate, and all mainstream platforms do. Without moderation, platforms are readily taken over by harassers and robots; and yet no moderation policy is value neutral. 

Flash points in the moderation debates include years-long protests over Facebook's policy of classifying (and later declassifying) breastfeeding photographs as ``obscene'' content; Facebook's controversial policy of taking down obscene but historically significant images, such as the Pulitzer Prize-winning ``Napalm Girl'' photograph notable for its role in bending public opinion on the Vietnam War; and, following the January 6 Capitol Hill riots, the wave of account suspensions that swept across Twitter, Facebook, Amazon, and even Pinterest. 

In all of these cases, platforms faced consequences in the marketplace, as well as brand-management challenges. From a legal standpoint, however, their autonomy has seldom been challenged. 

In the end, Gillespie provokes his readers to reconsider whether platforms should be entrusted with decisions that are inevitably political and affect all of us. Analyzing platforms through the lens of moderation raises fundamental questions about the sufficiency of current regulations. The moderation lens, however, seldom forces us to question the very validity of the intermediary-creator distinction. 

\subsection*{WHAT IS CONTENT CREATION, ANYWAY?} 
This article argues that major changes in both the technology used to curate content and the nature of user content itself are rapidly eroding the boundary between intermediaries and creators. 

First, breakthroughs in machine-learning algorithms and systems for intelligently assembling the underlying content into curated experiences have given companies the power to determine with unprecedented control not only what \emph{can be seen}, but also what \emph{will actually be seen} by users in service of whatever metric a company believes serves its business objectives. 

Second, unlike traditional bulletin board sites for sharing links to entire articles, or blogging platforms for sharing article-length musings, modern social media giants such as Facebook and Twitter traffic primarily (and increasingly) in microcontent—isolated snippets of text and photographs floating \emph{a la carte} through their ecosystems. 

Third, the largest platforms operate on such an enormous scale that their content contains nearly any assertion of fact (true or false), nearly any normative assertion (however extreme), and nearly any photograph (real or fake) floating through the zeitgeist. 

Platforms now enjoy vast expressive power to create media products for their users, limited only by the available atomic content and by the power of their algorithms, both of which are advancing rapidly because of economies of scale and advances in technology, respectively. 

We are not the first to suggest that curation fundamentally alters the distinction between platforms and creators. In a recently proposed amendment to Section 230, motivated by more pragmatic regulatory concerns, U.S. Representatives Anna G. Eshoo (D-Cal.) and Tom Malinowski (D-N.J.) recently proposed to reclassify those ``interactive computer service[s]'' (platforms) that ``used an algorithm, model, other computational process to rank, order, promote, recommend, amplify, or similarly alter the delivery or display of information'' as an ``information content provider'' (creator)~\citep{sec230eshoo}.
To be clear that the interpretation of these legal terms is faithful to the original meaning in Section 230, here is the official definition:

\begin{displayquote}
\emph{
The term information content provider means any person or entity that is responsible, in whole or in part, for the creation or development of information provided through the Internet or any other interactive computer service. 
} 
\end{displayquote}

Immediate legal goals aside, why target (algorithmic) content curation? At first glance, it might seem absurd that by virtue of curating content, an Internet service should assume not only some measure of responsibility, but also the very same status, vis-a-vis liability, as the creators of the underlying content. This distinction, however, may not actually be so far-fetched. 

Similar debates have arisen in the arts. Who can claim responsibility for a pop song that heavily samples preexisting audio? Are the Beastie Boys the creators of \emph{Paul's Boutique}, or do the creators of the original snippets have a sole right to that distinction? Can Jasper Johns be considered the creator for his prints and collages that repackage and juxtapose previous works of art (by himself and others)? 

With such derived works, claims to creatorship, rights to the spoils, and liability need not be mutually exclusive. This precedent suggests at least one sphere of life where people appear to be comfortable with the idea that those who produce microcontent and those who assemble it into larger-scale works can share the designation of \emph{creator}. 

Of course, the line must be drawn somewhere. The DJ does not create the music in the same way that the Beastie Boys do. Art galleries do not create art in the same way that Jasper Johns does. Beneath the neat system of legal categories lies a messy spectrum of creative activities. 

\subsection*{WHEN DOES CURATION BECOME CREATION?} 
Returning to the activities of web platforms, let's consider two extremes on the curation-creation spectrum. First, let's consider the activities of a typical aggregator website such as the Drudge Report, whose content consists entirely of outbound links to full articles that exist elsewhere on the Internet. Arguably, Drudge plays the role of the DJ, creating something more like a playlist than a song. 

Now consider the typical online blogger or the typical overworked journalist of the online era offering commentary or synthesis but not original reporting. They scour the Internet for content, assembling words, phrases, whole quotes, and photographs, all of which could be found elsewhere, to produce an article or post. Most readers undoubtedly concur that this qualifies as creation. Indeed, it is creation in the same sense that Twitter and Facebook users are creators of the content they post. 

Now consider the middle ground, where someone fashions content by assembling neither whole articles nor individual words but instead individual sentences, drawn from the entirety of the Internet, stripped of their original context, and assembled to present any desired picture of the discourse surrounding any topic. 

Legal scholars and politicians can debate whether this middle ground warrants official categorization as creation versus curation. It's hard to deny, however, that these acts indeed constitute a spectrum and that the curator of sentences bears greater resemblance to the curator of words than does the curator of articles. 

Today's platforms have been creeping steadily along this spectrum. From the earliest days, when a comparatively puny reservoir of content was presented in reverse chronological order, to the modern era's black-box systems that power Twitter's and Facebook's news feeds, there is a shifting landscape of actors that look less and less like disinterested utilities happy to transport any content that shows up in its plumbing and more and more like active creators of a media product. 

To be sure, activity along this spectrum is not uniform, even within a single platform. Take Twitter, for example: While the default news feed is indeed customized according to an opaque process, the content consists mostly of recent posts by (or retweeted by) individuals whom you follow. On the other hand, Twitter's Explore screen bears a striking resemblance to the middle-ground curator of sentences. They both present a set of hot topics, each titled according to some unknown process, and they curate, from the (often) millions of tweets on a topic, a chosen set to represent the story. 

In an era where many journalistic articles appearing in traditional venues consist of curated sets of tweets loosely connected by narrative and interpretation, the line separating intermediary from creator has grown so thin as to suggest the possibility that a double standard is already at play. 

\subsection*{WHERE DO WE GO NEXT?} 
While the focus here is on actions that platforms take to present content, this is not the only way they influence the information a user consumes. Platforms like Twitter and Facebook regularly translate messages across languages. Image-sharing platforms, such as Instagram and Snapchat, apply algorithmic transformations to photographs. 
As technology advances, the murky line between curation and creation is likely to become less, not more, distinct. 

In the future, platforms might not only translate across languages, but also paraphrase across dialects~\citep{lewis2020pre} or provide content summaries~\citep{wang2019summarizing}. They may move past applying cute filters and render whole synthetic images to specification~\citep{koh2021text}. Perhaps to mollify users aghast at the toxicity of the web, Twitter and Facebook might offer features to render messages more polite~\citep{madaan2020politeness}.

Coming up with policies that balance the competing desiderata of corporate accountability, economic vibrancy, and individual rights to free speech is difficult. This article does not presume to champion a single point on the curation-creation spectrum as the one true cutoff. Nor does it purport to offer definitive guidance on the viability of a system predicated on such a distinction in the first place. 

Instead, the goal here is to elucidate that there is indeed a spectrum between curation and creation. Furthermore, technological advances provide platforms with a powerful, diverse, and growing set of tools with which to build products that exist in the gray area between ``interactive computer services'' and ``information content providers.'' 

Regulating this influential and growing sector of the Internet requires recognition of the essential gray-scale nature of the problem and that we eschew reductive regulatory frameworks that shoehorn all online actors into simplistic systems of categorization. 

At some point, the increasing influence that modern platforms wield over user experiences must be accompanied by greater responsibilities. It is hard to decide the precise point along the intermediary-creator spectrum at which platforms should assume liability. The bill proposed by Representatives Eshoo and Malinowski suggests that such a  point has already been reached. Surely, Facebook's legal team would disagree. What is clear, however, is that today's platforms play a growing role in creating media products and that any coherent regulatory framework must adapt to this reality. 

\bibliography{refs}
\bibliographystyle{plainnat}

\end{document}